\documentclass[12pt]{iopart}

\newcommand{\be}{\begin{equation}}
\newcommand{\ee}{\end{equation}}
\newcommand{\bq}{\begin{eqnarray}}
\newcommand{\eq}{\end{eqnarray}}

\newcommand{\1}{1\!\!1}

\newcommand{\ket}[1]{\left | \, #1 \right\rangle}

\usepackage{graphicx}

\begin{document}

\title[Revealing anyonic features in a toric code quantum simulation]
{Revealing anyonic features in a toric code quantum simulation}

\author{J. K. Pachos$^1$, W. Wieczorek$^{2,3}$, C. Schmid$^{2,3}$, N. Kiesel$^{2,3}$,
R. Pohlner$^{2,3}$ and H. Weinfurter$^{2,3}$}
\address{$^1$School of Physics \& Astronomy, University of Leeds,
Leeds LS2 9JT, UK,\\
$^2$Max-Planck-Institute of Quantum Optics, D-85748 Garching, Germany,\\
$^3$Department for Physics, Ludwig-Maximilians-Universit\"at M\"unchen,
D-80799 M\"unchen, Germany. }

\begin{abstract}

Anyons are quasiparticles in two-dimensional systems that show statistical
properties very distinct from those of bosons or fermions. While their
isolated observation has not yet been achieved, here we perform a quantum
simulation of anyons on the toric code model. By encoding the model in the
multi-partite entangled state of polarized photons, we are able to
demonstrate various manipulations of anyonic states and, in particular,
their characteristic fractional statistics.

\end{abstract}


\maketitle

\section{Introduction}

In three spatial dimensions, only two types of statistical behaviors have
been observed dividing particles into two groups: bosons and fermions. If
one is restricted to two-dimensional systems the situation changes. There,
anyons~\cite{Wilczek82} can appear, which exhibit fractional statistics that
ranges continuously from bosonic to fermionic. Anyons are responsible for
the fractional quantum Hall effect~\cite{Tsui82} and it has been
demonstrated that they could be realized as quasiparticles in highly
entangled many-body systems. Formally, the properties of anyons are
described by two-dimensional topological quantum field
theories~\cite{Witten} that dictate their trivial dynamical, but complex
statistical behavior. In general, it is expected that such topological
quantum field theories come into effect at low energies of highly correlated
many-body systems, such as quantum liquids~\cite{Anderson}. However, the
observation of anyonic features requires high population in the system's
ground state, high purity samples and, above all, the ability to separate
the anyonic effects from the dynamical background of the strongly
interacting system. In spite of significant experimental
progress~\cite{Goldman} the fractional statistics of anyons has not yet been
conclusively confirmed due to the complexity of these systems~\cite{Kim}.

To demonstrate characteristic anyonic features we employ here a different
strategy. We simulate the toric code model~\cite{Kitaev2003} in a quantum
system without the continuous presence of interactions~\cite{Han}. The toric
code is a two-dimensional topological lattice model, where anyons are
spatially well localized. Recently, several proposals showed how the toric
code could possibly be implemented on extended lattices of qubits enabling
one to employ anyonic states for, e.g., protected quantum
computation~\cite{Jiang08}. While such a quantum computational scheme
requires large systems, here we show that a number of anyonic properties can
be demonstrated by considering four-partite entangled
Greenberger-Horne-Zeilinger (GHZ) states~\cite{GHZ}. These states are
dynamically encoded in the polarization of photons, rather than produced via
cooling~\cite{Han}. This provides the additional advantage of being subject
to negligible decoherence due to the weak coupling of photons to the
environment. Polarized photons as qubits are a well understood and
controllable quantum optical system that already allowed to observe the
four-qubit GHZ state~\cite{Pan01}. Here we employ this system to simulate
anyonic behavior.

\section{Toric code model}

\subsection{Hamiltonian and ground state}

\begin{figure}[t!h]
\begin{center}
{\includegraphics{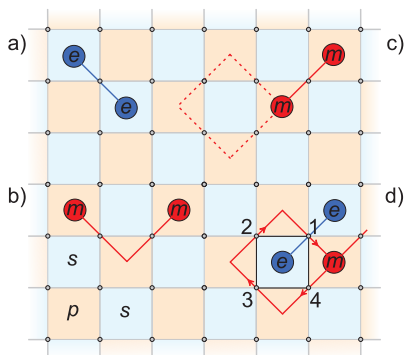}}
\caption{The toric code lattice with qubits at the vertices of a square lattice
with two types of plaquettes. Light blue and red plaquettes correspond to
the s- and p-plaquettes, respectively. Qubit rotations enable manipulations
of anyons on neighboring plaquettes. (a) Application of $\sigma^z$ on a
single qubit yields two $e$-type anyons placed at the neighboring $s$
plaquettes, where the string passes trough the rotated qubits. Similarly,
$m$ anyons are created on $p$ plaquettes by $\sigma^x$ rotations. (b) Two
$\sigma^x$ rotations create two pairs of $m$-type anyons. If one anyon from
each pair is positioned on the same plaquette then they annihilate, thereby
connecting their strings. (c) When a part of a string forms a loop around
unpopulated plaquettes, the loop cancels (dashed). (d) Here we restrict to
one $s$ plaquette and four neighboring $p$ plaquettes. Anyon $e$ is produced
by a $\sigma^z$ on qubit 1 (blue), $|e\rangle =\sigma^z_1|\xi\rangle$. This
system can support the circulation of an $m$ around an $e$
anyon.\label{fig-1}}
\end{center}
\end{figure}

The anyonic model under consideration is based on the toric code proposed by
Kitaev~\cite{Kitaev2003} and is described in detail in tutorial
introductions~\cite{Preskill}. This system can be defined on a
two-dimensional square lattice with interacting qubits placed at its
vertices. To facilitate the description of the interactions let us split the
lattice into two alternating types of plaquettes labelled by $p$ and $s$, as
in \fref{fig-1}. The defining Hamiltonian is
\begin{equation}
\mathcal{H}= -\sum_p \sigma^z_{p,1} \sigma^z_{p,2} \sigma^z_{p,3}
\sigma^z_{p,4} -\sum_s \sigma^x_{s,1} \sigma^x_{s,2} \sigma^x_{s,3}
\sigma^x_{s,4}~, \label{Ham1}
\end{equation}
where the summations run over the corresponding plaquettes and the indices
$1,...,4$ of the Pauli operators, $\sigma^z$ and $\sigma^x$, enumerate the
vertices of each plaquette in a counter-clockwise fashion. Each of the $s$-
or $p$-plaquette interaction terms commute with the Hamiltonian as well as
with each other. Thus, the model is exactly solvable and its ground state is
explicitly given by~\cite{Frank}
\begin{equation}
|\xi\rangle = \prod_s{\frac{1}{\sqrt{2}}}\big(\1+\sigma^x_{s,1}
\sigma^x_{s,2} \sigma^x_{s,3} \sigma^x_{s,4}\big)|00...0\rangle ~,
\label{state}
\end{equation}
with $\sigma^z|0\rangle=|0\rangle$. The state $|\xi\rangle$ represents the
anyonic vacuum state and it is unique for systems with open boundary
conditions.

\subsection{Anyonic quasiparticles}

Starting from this ground state one can excite pairs of anyons connected by
a string on the lattice using single qubit operations. More specifically, by
applying $\sigma^z$ on some qubit of the lattice a pair of so called
$e$-type anyons  is created on the two neighboring $s$ plaquettes
[\fref{fig-1}(a)] and the system is described by the state $|e\rangle
=\sigma^z |\xi\rangle$. An $m$ pair of anyons lives on the $p$ plaquettes
and is obtained by a $\sigma^x$ operation. The combination of both creates
the composite quasiparticle $\epsilon$ with $|\epsilon\rangle
=\sigma^z\sigma^x|\xi\rangle= i\sigma^y|\xi\rangle$. Two equal Pauli
rotations applied on qubits of the same plaquette create two anyons on this
plaquette. The fusion rules ($1\times1 = e\times e=m\times m
=\epsilon\times\epsilon=1$, $e\times m =
\epsilon$, $1\times e = e$, etc., where $1$ is the vacuum state~\cite{Preskill})
describe the outcome from combining two anyons. In the above example, if two
anyons are created on the same plaquette then they annihilate. This
operation also glues two single strings of the same type together to form a
new string, again with a pair of anyons at its ends [Figure~\ref{fig-1}(b)].
If the string forms a loop, the anyons at its end annihilate each other,
thus removing the anyonic excitation. In case that only a part of the string
forms a loop, the string gets truncated [Figure~\ref{fig-1}(c)]. For
non-compact finite systems, such as the one we consider here, a string may
end up at the boundary describing a single anyon at its free endpoint.

\subsection{Anyonic statistics}

Anyonic statistics is revealed as a non-trivial phase factor acquired by the
wave function of the lattice system after braiding anyons, i.e., after
moving an $m$ anyon around an $e$ anyon [Figure~\ref{fig-1}(d)] or vice
versa. Consider the initial state
$|\Psi_{\rm{ini}}\rangle=\sigma^z_1|\xi\rangle =|e\rangle$. If an anyon of
type $m$ is assumed to be at a neighboring $p$ plaquette it can be moved
around $e$ along the path generated by successive applications of $\sigma^x$
rotations on the four qubits of the $s$ plaquette. The final state is
\begin{equation}
|\Psi_{\rm{fin}}\rangle = \sigma^x_1 \sigma^x_2 \sigma^x_3
\sigma^x_4 |\Psi_{\rm{ini}}\rangle =-\sigma^z_1(\sigma^x_1\sigma^x_2
\sigma^x_3\sigma^x_4 |\xi\rangle) = -|\Psi_{\rm{ini}}\rangle ~.
\label{stat}
\end{equation}
Such a minimal loop, which vanishes the moment it is closed, is analogous to
the application of the respective interaction term
$C_s=\sigma^x_{s,1}\sigma^x_{s,2}\sigma^x_{s,3}\sigma^x_{s,4}$ (or
$C_p=\sigma^z_{p,1}\sigma^z_{p,2}\sigma^z_{p,3}\sigma^z_{p,4}$) of the
Hamiltonian. This operator has eigenvalue $+1$ for all plaquettes of the
ground state $|\xi\rangle$, whereas it signals an excitation, e.g.,
$|\Psi_{\rm{ini}}\rangle$, with eigenvalue $-1$, when applied to the
plaquette where an anyon resides. However, \eref{stat} is much more general,
as the actual path of the loop is irrelevant. It is the topological phase
factor of $-1$, which reveals the presence of the enclosed anyon.
Alternatively, we can interpret \eref{stat} as a description of twisting
$\epsilon$, the combination of an $e$ and an $m$-type anyon, by $2\pi$. The
phase factor of $-1$ thereby reveals its $4\pi$-symmetry, characteristic for
half spin, fermionic particles~\cite{4PI}. Note that the $e$ ($m$) anyons
exhibit bosonic statistics with respect to themselves~\cite{Kitaev2003}.

\section{Experimental implementation}

\subsection{Minimal instance of the toric code model}

As this two-dimensional system is well suited for demonstrating
characteristic anyonic features, the question arises how big the lattice has
to be. It turns out that four qubits of a single $s$ plaquette represent the
minimal unit of the toric code model~\cite{Pachos2}. In this case, the four
neighboring $p$ plaquettes are represented only by their corresponding
links. Hence, while the presence of an $e$ anyon is detected by the
plaquette operator $C_s$, the presence of $m$ anyons is detected by the
eigenvalue of the four $C_p'=\sigma^z_{p,i}\sigma^z_{p,j}$ operators that
correspond to the adjacent links of the $p$ plaquettes. Consequently, the
product determining the ground state [Eqn. (\ref{state})] reduces to only
one factor for the single $s$-plaquette resulting in a four qubit GHZ state,
$\ket{\xi} = (\ket{0000}+\ket{1111})/\sqrt{2}$. This is an eigenstate of the
relevant $C_s$ and $C_p'$ operators with eigenvalue $+1$. An eigenvalue $-1$
for any of these operators denotes the presence of the corresponding anyon.


\begin{figure}[t!h]
\begin{center}
{\includegraphics{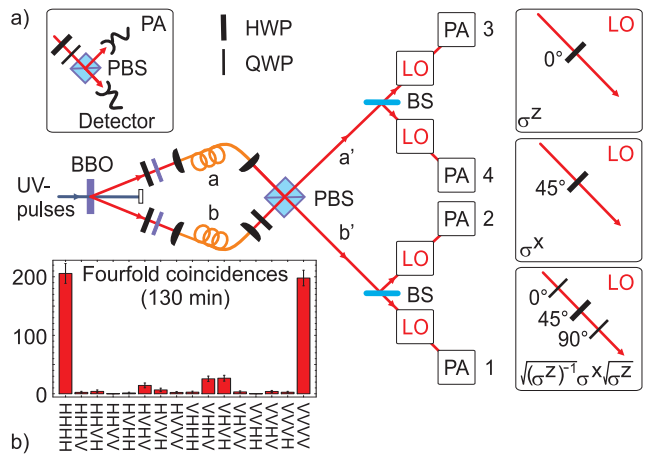}}
\caption{(a) The experimental set-up for the observation of the anyonic
vacuum state of the toric code model, the demonstration of anyonic features
and the verification of anyonic statistics. The photons are created in a
2\,mm thick $\beta$ Barium Borate (BBO) crystal, which is pumped by
femtosecond ultraviolet (UV) pulses. Walk-off effects are compensated for by
a HWP and a 1\,mm thick BBO crystal. For the observation of the anyonic
states, we are interested in the second order emission of the
SPDC~\cite{Weinfurter}, initially emitted in the two spatial modes $a$ and
$b$: $\propto |2{\rm{H}}\rangle_a|2{\rm{V}}\rangle_b
+|2{\rm{V}}\rangle_a|2{\rm{H}}\rangle_b+|{\rm{HV}}\rangle_a
|{\rm{HV}}\rangle_b$. The photons are further processed by a HWP in mode
$b$, which transforms $|\rm{H}\rangle$ into $|+\rangle$ and $|\rm{V}\rangle$
into $|-\rangle$ [with
$|\pm\rangle=(|\rm{H}\rangle\pm|\rm{V}\rangle)/\sqrt{2}$]. The state
$|\xi\rangle$ is observed behind the PBS and the BSs under the condition of
detecting a photon in each of the four modes 1, 2, 3, 4 (average count rate:
2.8 min$^{-1}$). The local operations (LO) for the demonstration of anyonic
features are implemented by operation specific HWPs and quarter-wave plates
(QWPs). The polarization state of each photon is analyzed (PA) with a HWP
and a QWP in front of a PBS. (b) Measurement outcomes of $|\xi\rangle$ in
the basis $\sigma^z_i$ for each qubit.
\label{fig-2}}
\end{center}
\end{figure}

\subsection{State implementation}

In our simulation the qubits supporting the anyonic states are encoded in
the polarization of single photons propagating in well-defined spatial
modes. This means that for the anyonic vacuum and for all obtained final
states our system is of the form $|\rm{GHZ}^{\phi}\rangle=(|H_1 H_2 H_3
H_4\rangle+e^{i\phi}|V_1 V_2 V_3 V_4\rangle)/\sqrt{2}$. The indices (omitted
in the following) label the spatial mode of each photon, i.e., they
represent the position of the qubit as in Figure~\ref{fig-1}(d), and H (V)
denote linear horizontal (vertical) polarization, representing a logical $0$
($1$). To obtain such four-photon entangled states, the second order
emission of a non-collinear type-II spontaneous parametric down conversion
process (SPDC)
\cite{Kwiat}, yielding four photons in two spatial modes, is overlapped on a
polarizing beam splitter (PBS) and afterwards symmetrically split up into
four spatial modes by two polarization independent beam splitters (BS), see
Figure~\ref{fig-2}(a) \cite{wieczorek}. Prior to the second order
interference at the PBS the polarization of two of the photons is rotated by
a half-wave plate (HWP). Under the condition of detecting one photon in each
spatial mode we observe the desired states.

As we are interested in the vacuum $|\xi\rangle=|\rm{GHZ}^{0}\rangle$ and
the anyonic state $|e\rangle=\sigma_i^z|\xi\rangle=|\rm{GHZ}^{\pi}\rangle$,
the successful simulation and demonstration of anyonic features relies on a
careful distinction and characterization of these two orthogonal GHZ states.
This comprises the confirmation of genuine four-partite entanglement and a
method to reveal the phase $\phi$. For this purpose, state analysis is
performed by measuring the correlation, $c_z$, in the $\sigma^z_i$ basis for
each qubit and the correlation function
$c_{xy}(\gamma)=\sigma_1(\gamma)\otimes\sigma_2(\gamma)\otimes\sigma_3
(\gamma)\otimes\sigma_4(\gamma)$ with
$\sigma_i(\gamma)=[(\cos{\gamma})\sigma^y_i + (\sin{\gamma})\sigma^x_i]$
\cite{Sacket}. This measurement is ideally suited to prove the coherent
superposition of the terms $|\rm{H H H H}\rangle$ and $|\rm{V V V V}\rangle$
and to determine their relative phase $\phi$. It is obtained from a
measurement of all qubits along the same direction in the $x$-$y$ plane of
the Bloch sphere. For a GHZ$^\phi$ state its expectation value shows an
oscillation, $\langle c_{xy}(\gamma)\rangle
=\mathcal{V}\cos{(4\gamma+\phi)}$, with a period of one quarter of $2\pi$.
This is a unique signature of four qubit GHZ entanglement. The visibility of
this oscillation $\mathcal{V}$ is 1 for pure GHZ states. For a general state
it equals twice the element
$\rho_{\rm{HHHH,VVVV}}=\rm{Tr}(|\rm{HHHH}\rangle\langle
\rm{VVVV}|\rho)$ of the corresponding density matrix $\rho$.

To analyze the performance of the set-up, first, the observation of the
state $|\xi\rangle=|\rm{GHZ}^0\rangle$ is confirmed. Figure~\ref{fig-2}(b)
shows all possible measurement outcomes in the
$\sigma^z_1\sigma^z_2\sigma^z_3\sigma^z_4$ basis, where the expected
populations of $|\rm{H H H H}\rangle$ and $|\rm{V V V V}\rangle$ are clearly
predominant with probabilities of $P_{\rm{HHHH}}=(41.2\pm3.4)\%$ and
$P_{\rm{VVVV}}=(39.6\pm2.7)\%$, respectively, close to the expected
$P_{\rm{HHHH}}=P_{\rm{VVVV}}=50\%$. We find a correlation of $\langle
c_{z}\rangle =0.908\pm0.047$. The population of other terms is caused by
noise arising from higher-order emissions of the down conversion and a
remaining degree of distinguishability of photons at the PBS. The dependence
of $\langle c_{xy}(\gamma)\rangle$ for the state $|\xi\rangle$ is displayed
in Figure~\ref{fig-3}(a) from which we infer a visibility $\mathcal{V}$ of
$(68.3\pm1.1)\%$. This value is obtained from a weighted least squares fit
to a Fourier decomposition of $\langle c_{xy}(\gamma)\rangle$ considering
only even components up to order 4: $\langle
c_{xy}(\gamma)\rangle=\sum_{k=0}^2a_k\cos(2k\cdot\gamma)+b_k\sin(2k\cdot\gamma)$.
Only these components can originate from physical states.  From the phase of
the correlation function we deduce the phase of the GHZ$^0$ state to be
$\phi=(0.02\pm0.01)\pi$, close to the expected value of $0$.

The visibility $\mathcal{V}$ together with the populations $P_{\rm{HHHH}}$
and $P_{\rm{VVVV}}$ allow further to determine the fidelity
$F=\langle\rm{GHZ}|\rho|\rm{GHZ}\rangle=(\mathcal{V}+P_{\rm{HHHH}}+
P_{\rm{VVVV}})/2$. We obtain a value of $(74.5\pm2.2)\%$ (for measurement
data see also Table~\ref{Table1}). The fidelity is not only important for an
estimation of state quality but can also be applied to verify genuine
four-partite entanglement, an essential element in the presented toric code
model. With a proper entanglement witness a fidelity greater than 50\% is
sufficient~\cite{Tot05}. The experimentally observed fidelity is clearly
above this bound, i.e., it is of high enough quality to allow for the
demonstration of the anyonic properties.

\subsection{Anyonic manipulations and detection of anyonic statistics}

\begin{figure}[t!h]
\begin{center}
{\includegraphics{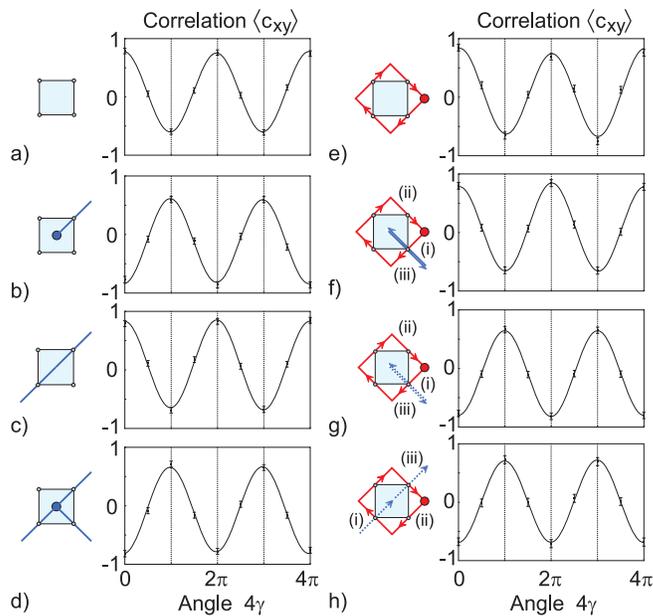}}
\caption{Pictorial representations of various evolutions of anyons, together
with the correlation function $\langle c_{xy}\rangle$. The phase of $\langle
c_{xy}\rangle$ directly gives the phase $\phi$ of $\ket{{\rm GHZ}^\phi}$.
(a) The vacuum state, $|\xi\rangle$. (b) Generating an $e$-type anyonic
state $|e\rangle=|{\rm GHZ}^{\pi}\rangle$ by a $\sigma^z$ operator. (c) A
second $\sigma^z$ removes the anyon from the plaquette and elongates the
string resulting in the GHZ$^0$ state. (d) A third $\sigma^z$ creates
another pair of anyons, connected with the string already traversing the
plaquette. (e) Moving an $m$ anyon around an empty $s$ plaquette or (f)
first creating an $e$ anyon on the plaquette (i), performing the loop (ii),
and removing the $e$ anyon from the plaquette again (iii) give identical
final states up to a global phase factor. (g) The global phase, $\pi$, is
revealed by superimposing the two above evolutions, thereby clearly proving
anyonic statistics. (h) An alternative path for the anyon $e$ that gives the
same fractional statistical phase, $\pi$.
\label{fig-3}}
\end{center}
\end{figure}

We start by analyzing the state change under the creation (and annihilation)
of anyons, thereby demonstrating the characteristic fusion rules. Applying
$\sigma_1^z$ in mode $1$ creates an $e$-type anyon on the $s$ plaquette
resulting in a GHZ$^{\pi}$ state, Figure~\ref{fig-3}(b). This is clearly
proven by the phase $\phi= 1.02 \pi$ of the correlation function. A further
application of $\sigma_j^z$ on any other mode, say $j=3$, changes the state
of the plaquette according to $e \times e = 1$, and the anyon is moved away
from the $s$ plaquette under investigation. These two $\sigma^z$ rotations
represent a string connecting two anyons, which traverses the particular
plaquette without influencing its state, as confirmed by the observation of
the initial GHZ$^0$ state [$\phi=0.01\pi$, Figure~\ref{fig-3}(c)]. A further
$\sigma_j^z$ rotation on one of the remaining vertices, e.g., $j=4$, creates
an $e$ occupation on the plaquette. Thus, the observation of GHZ$^\pi$
[$\phi=1.03\pi$, Figure~\ref{fig-3}(d)] demonstrates the non-trivial version
of the $1\times e =e$ fusion rule. Alternatively, this can be seen as
demonstrating the invariance of an anyonic state, when a string traverses
the plaquette.

To detect the major feature of anyons, their non-trivial statistical phase
acquired when moving one anyon around the other, we employed an interference
measurement, which makes their \emph{overall} phase factor visible [see
\eref{stat} and Figure~\ref{fig-1}(d)]. Let us study the two evolutions
separately. The first is to create a pair of two $m$ anyons, move one of
them around the empty $s$ plaquette and then annihilate them. This evolution
is equivalent to having a string ending up at a $p$ plaquette with an $m$
anyon at its endpoint and circulating that anyon around the $s$ plaquette.
The obtained loop operator $C_s$, obviously, does not change the vacuum
state and, thus, results in GHZ$^0$ [$\phi=-0.02\pi$,
Figure~\ref{fig-3}(e)]. Alternatively, we (i) create an $e$ anyon on the
plaquette, (ii) encompass it with the loop of an $m$ anyon, and (iii) remove
it again [Figure~\ref{fig-3}(f)]. The whole evolution is, in analogy to
\eref{stat}, described by $\sigma^z_4(\sigma^x_1\sigma^x_2
\sigma^x_3\sigma^x_4)\sigma^z_4
|\xi\rangle=-|\xi\rangle=-|\rm{GHZ}^0\rangle$. The correlation function
determines faithfully the angle $\phi=-0.01\pi$ [Figure~\ref{fig-3}(f)], but
is blind to the characteristic overall phase factor. This, finally, can be
observed by (i) first generating the {\em superposition}
$\mathrm{e}^{-i\pi/4}(|\xi\rangle+i|e\rangle)/\sqrt{2}$ of having an anyon
$e$ on the plaquette or not by the unitary operation
$({\sigma_i^z})^{-1/2}$; (ii) moving the $m$ anyon around this superposition
gives $\mathrm{e}^{-i\pi/4}(|\xi\rangle-i|e\rangle)/\sqrt{2}$, the
superposition of the above evolutions; and (iii) the application of the
inverse rotation $({\sigma_i^z})^{1/2}$ makes the phase difference visible
resulting in $ -i|e\rangle=-i|\rm{GHZ}^\pi\rangle$. The observed value of
$\phi=1.00\pi$ [Figure~\ref{fig-3}(g)] clearly proves the phase acquired by
the braiding and therefore the anyonic statistics~\cite{Comment2}. To
reinforce the observation of the obtained fractional phase, we perform an
additional interference experiment with an alternative path. Now, the
superposition of an $e$ anyonic state and the ground state is generated in
the $s$-plaquette by rotating one qubit and it is removed by rotating
another qubit, as seen in~Figure~\ref{fig-3}(h). If we move an $m$ anyon
around the plaquette in between the two steps, we observe {\em again} the
characteristic phase shift of $-1$.

\begin{table}[h]
\begin{center}
\begin{tabular}{lcccc}
                        \hline\hline
            &Phase $\phi$&Fidelity $F$\\ \hline
            \multicolumn{3}{l}{Vacuum state}\\
            $|\xi\rangle$&$(0.02\pm0.01)\cdot \pi$&$(74.5\pm2.2)\%$\\
            \hline
            \multicolumn{3}{l}{Creation of a single $e$-type anyon}\\
            $\sigma^z_1|\xi\rangle$&$(1.02\pm0.01)\cdot\pi$&$(74.9\pm2.8)\%$\\
            $\sigma^z_2|\xi\rangle$&$(1.00\pm0.01)\cdot\pi$&$(74.2\pm2.7)\%$\\
            $\sigma^z_3|\xi\rangle$&$(1.01\pm0.01)\cdot\pi$&$(76.5\pm3.2)\%$\\
            $\sigma^z_4|\xi\rangle$&$(0.97\pm0.02)\cdot\pi$&$(76.2\pm3.7)\%$\\
            \hline
            \multicolumn{3}{l}{String passing through the $s$ plaquette}\\
            $\sigma^z_1\sigma^z_2|\xi\rangle$&$(0.02\pm0.01)\cdot \pi$&$(77.4\pm2.8)\%$\\
            $\sigma^z_1\sigma^z_3|\xi\rangle$&$(0.01\pm0.01)\cdot \pi$&$(77.3\pm2.5)\%$\\
            $\sigma^z_1\sigma^z_4|\xi\rangle$&$(-0.01\pm0.01)\cdot \pi$&$(76.3\pm2.6)\%$\\
            \hline
            \multicolumn{3}{l}{String passing through the $s$ plaquette populated with an anyon}\\
            $\sigma^z_2\sigma^z_4|e\rangle$&$(1.02\pm0.01)\cdot\pi$&$(75.2\pm2.4)\%$\\
            $\sigma^z_3\sigma^z_4|e\rangle$&$(1.03\pm0.01)\cdot\pi$&$(76.7\pm2.7)\%$\\
            $\sigma^z_1\sigma^z_4|e\rangle$&$(1.02\pm0.02)\cdot\pi$&$(74.5\pm3.1)\%$\\
            \hline
            \multicolumn{3}{l}{Loop around an unpopulated $s$ plaquette}\\
            $C_s|\xi\rangle$&$(-0.02\pm0.02)\cdot \pi$&$(76.6\pm3.4)\%$\\
            \hline
            \multicolumn{3}{l}{Loop around a populated $s$ plaquette followed by annihilation of the anyon}\\
            $(\sigma^z_4)C_s(\sigma^z_4)|\xi\rangle$&$(-0.01\pm0.01)\cdot\pi$&$(76.8\pm2.8)\%$\\
            \hline
            \multicolumn{3}{l}{Interference procedure to reveal anyonic statistics}\\
            $(\sigma^z_1)^{1/2}C_s(\sigma^z_1)^{-1/2}|\xi\rangle$&$(1.03\pm0.01)\cdot\pi$&$(76.1\pm3.0)\%$\\
            $(\sigma^z_2)^{1/2}C_s(\sigma^z_2)^{-1/2}|\xi\rangle$&$(0.99\pm0.01)\cdot\pi$&$(73.5\pm3.6)\%$\\
            $(\sigma^z_3)^{1/2}C_s(\sigma^z_3)^{-1/2}|\xi\rangle$&$(1.01\pm0.01)\cdot\pi$&$(75.2\pm3.0)\%$\\
            $(\sigma^z_4)^{1/2}C_s(\sigma^z_4)^{-1/2}|\xi\rangle$&$(1.00\pm0.01)\cdot\pi$&$(75.8\pm2.5)\%$\\
            $(\sigma^z_1)^{1/2}C_s(\sigma^z_3)^{-1/2}|\xi\rangle$&$(1.00\pm0.01)\cdot\pi$&$(78.7\pm3.2)\%$\\
            \hline\hline
\end{tabular}
\end{center}
\caption{Experimental measurement data of the anyonic states.}\label{Table1}
\end{table}

\section{Conclusions}

Our results show that we can create, manipulate and detect the toric code
states by encoding them in a simple physical system. In our quantum
simulation we demonstrated several of the fusion rules and we provided
supporting evidence for the fractional statistics of the toric code anyons.
Multiqubit toric code states are known to be useful for novel types of
quantum error correction~\cite{Kitaev97}. Extending our experimental results
presented here to larger, scalable quantum systems~\cite{Robert,Zoller} will
enable the application of the toric code for quantum information processing
in the future~\cite{Freedman}.


Note added: With the completion of this work we became aware of a similar
experimental implementation, closely resembling \cite{Han}, that was
simultaneously performed \cite{Pan07Ay}.

\ack

We would like to thank Phillipe Grangier, Peter Zoller and Robert
Raussendorf for inspiring conversations. We acknowledge the support of this
work by EPSRC, the Royal Society, the DFG-Cluster of Excellence MAP and the
EU Projects QAP, EMALI and SCALA. W.W.~acknowledges support by QCCC of the
Elite Network of Bavaria and the Studienstiftung des dt.~Volkes.


\begin{thebibliography}{99}
\bibitem{Comment2}
Note that during the braiding evolution no dynamical phases are generated as
there is no Hamiltonian present and the Pauli rotations are performed with
zero order wave plates, thus, adding insignificant erroneous phases.
\end{thebibliography}
\end{document}